# Rheothermodynamics of transient networks

Jean-François Palierne

*Laboratoire de Physique/URA 1325 du CNRS, Ecole Normale Superieure de Lyon,*
*46 allée d'Italie/ 69364 Lyon Cedex 07, France*

Synopsis :   The transient network model of Green-Tobolsky [1946], Yamamoto [1956] and Tanaka-Edwards [1992] is formulated within the frame of thermodynamics of irreversible processes, using as a fundamental quantity the chemical potential associated to the connection of strands to the network and treating these connections as chemical-like reactions. All thermodynamic quantities are thus naturally defined in and out of equilibrium. Constitutive equations are derived, giving the stress and the heat production as functions of the thermomechanical history. The Clausius-Duhem inequality, stating that the source of entropy is non-negative, is shown to hold for any thermomechanical history, ensuring the thermodynamic consistency of our model. The presented model includes the Green-Tobolsky model, whereas those of Yamamoto and Tanaka-Edwards fit within ours on the condition that their free parameters obey a detailed balance condition stemming form Boltzmann equilibrium statistics.

**Introduction**

The transient network models first aimed at a molecular description of the viscoelasticity of polymers in the entangled regime, either in the melt state or in concentrated solutions. These systems exhibit a rubberlike elasticity when subjected to fast deformations but, in contrast with rubber, their stress relaxes when the deformation is kept constant. The pioneering work of Green and Tobolsky (1946) started the subject by suggesting that the laws of rubber elasticity were valid in non-crosslinked polymeric systems, the rubber crosslinks being replaced by the entanglements between chains. An important characteristics of entanglements is that Brownian motion enables the chains them to disentangle and reentangle with other chains, in contrast with the permanent character of the rubber chemical crosslinks. This picture of entangled systems has been since modified by the reptation theory, but a more definite embodiment of transient networks is provided by associating polymers, which carry sticking sites able to form transient bounds with one another. There exists a great variety of binding mechanism, such as solvent complexation, local crystallisation or hydrogen bonding, to name just a few. Recent interest has been focused on water solutions of end-capped associating thickeners, obtained by grafting hydrophobic paraffinic groups at the ends of a water-soluble polymer chain. In solution, these strands join their end groups in tiny micelles acting as the connection points of a three-dimensional network. Brownian agitation and/or mechanical action can force an end group out of a micelle and, after having diffused in water, this end group will eventually join another micelle. The elastic links between micelles are provided by the hydrophilic chains, which can be made monodisperse, making this system a well-characterised and simple transient network [Annable et al (1992)].

All transient networks share the following characteristics: They are made of elastic strands joined by temporary, reversible junctions. The creation or the disappearance of junctions causes the strands to connect to or to disconnect from the network. The junctions' temporary character causes the network to be continuously reworked, some connected strands disconnecting while free ones connect. When the sample is submitted to deformation, the disappearing network has been strained by the deformation accumulated during its lifetime, while the newly created one incorporates strands that carry less or no stress, having had time to relax before their connection. The stress relaxation thus directly originates in the transient character of the network, and the relaxation time is related to the junction lifetime.

Such simple characteristics lend themselves to a molecular description of both the thermodynamics and the viscoelasticity of transient networks. The first attempt is the Green-Tobolsky model, which considers a network made of Gaussian strands which are created at a constant rate according to an equilibrium distribution, and have a constant probability per unit time to disappear, i.e. to disconnect from the network. As a consequence of these assumptions, Green and Tobolsky (1946) showed that the stress is related to the strain history according to a codeformational Maxwell equation, and that isothermal periodic closed-cycle deformations lead to dissipation of mechanical energy. The codeformational Maxwellian stress-strain relation however suffers from limitations: the simple shear steady-state viscosity shows no shear-thinning, the second normal stress difference vanishes, and the elongational viscosity diverges at a critical extension rate. Yamamoto (1956-1957-1958) addressed these difficulties by introducing a strand disconnection rate that increases when the strand is extended, and by relaxing the Gaussian strand assumption. He was able to work out an expression for the instantaneous energy dissipation. This model better agrees with experiment when a moderate dependence of the disconnection rate versus the strand extension is assumed, but when pushed too far, it can predict thermodynamically impossible behaviours. Yamamoto himself pointed out in his 1957 paper that the viscosity can be made negative by choosing a disconnection rate proportional to $n$-th power of the strand extension, with $n>5$. The same difficulties arise in the Tanaka-Edwards model (1992a-d), which adds to the preceding one a balance between connected and disconnected strands, making the connection rate proportional to the number of disconnected strands.

Our aim is to formulate the simplest complete rheothermodynamic model of transient networks including Yamamoto and Tanaka-Edwards' ideas. We call « Rheothermodynamics » the thermodynamics of systems submitted to deformation, and « complete » means that all thermodynamic quantities, including the stress and the rate of heat production as well as the thermodynamic potentials, can be related to the thermomechanical history of the system, i.e. the past and present temperature and deformation. The model is kept as simple as possible by minimising the number of assumptions, with no recourse to special mechanisms which would only operate out of equilibrium. Since most thermomechanical histories drive the system out of equilibrium, our thermodynamic description must be framed within the non-equilibrium thermodynamics of irreversible processes. The present paper comprises seven sections arranged in order of increasing specialisation. The first section formulates the general thermodynamic relations who do not depend on the precise form of the relaxation phenomena, using as fundamental quantities the strand distribution function and its thermodynamically conjugate quantity, the strand chemical potential. The second principle is formulated in terms of the Clausius-Duhem inequality. The second section specialises to ideal systems, corresponding to non-interacting strands, and the third one examines the constraints the Boltzmann equilibrium statistics places on the model. The fourth section introduces the kinetic processes, namely the connection/disconnection reactions and the heat conduction, allowing to formulate the evolution equation for the distribution function, thus completing the rheothermodynamic model this paper aims at. The model is then shown to be thermodynamically consistent for any given thermomechanical history. The special case of entropic systems, such as polymers, is treated in the fifth section. The sixth section is devoted to the limiting case of a connection rate independent of the number of connected strands. Eventually, the seventh section considers some restricting assumptions usually made in the transient network literature and examines how the formerly published models fit within ours, uncovering some thermodynamic inconsistencies.

## 1/ General

The system we consider is made of strands, in solution or in the melt, able to form a network through mutual transient connections. When connected, strands join two different points of the network separated by a vector $\vec{h}$ we call the connector. The connection to the network as well as the disconnection are instantaneous events, and the connection/disconnection rate is supposed to be slow enough for the kinetic degrees of freedom to equilibrate within this time scale. The network is constrained to deform affinely to the macroscopic deformation, whereas the free strands are assumed to undergo a Brownian diffusion in the $\vec{h}$ space. The fundamental point of this paper is that the connection/disconnections can be treated as chemical-like reactions between free and connected strands, and the motion in the $\vec{h}$ space can be considered as well as a continuous chemical reaction in the sense of Prigogine and Mazur (1953). With such provisions, transient networks belong to the field of thermodynamics of irreversible processes in systems with internal degrees of freedom [Prigogine and Mazur (1953), de Groot and Mazur (1984) § X-6]. The thermodynamic functions are thereby defined in non-equilibrium situations exactly as in a system undergoing a chemical reaction. For the sake of simplicity, we consider only one kind of strands ; the generalisation to a distribution of strand species can be made along the line of Lodge (1968), by adding the contributions of the various species to the various thermodynamic potentials. In the same spirit of simplicity, the strands we consider have only two possible connection states : they can be either free or connected, therefore dangling strands are ignored.

Consider a sample of volume $V$, containing $N$ strands and $N_S$ solvent molecules. This sample is taken small enough for space-dependent quantities such as the stress, strain and temperature to be considered uniform within $V$. The strands are characterised by a connection state $\alpha$ taking one of the two possible values $F$, for *free*, and $C$, for *connected to the network*, and by a connector $\vec{h}$ with the dimension of a length, belonging to the three-dimensional space tangent to the physical space. The strand distribution function $\Gamma_{\alpha\vec{h}}$ is such that $\Gamma_{\alpha\vec{h}} d^3\vec{h}$ is the number of strands within volume $V$, in connection state $\alpha$, and with vector $\vec{h}$ within the differential volume element $d^3\vec{h} = dh_1 dh_2 dh_3$. $\Gamma_{\alpha\vec{h}}$ has the dimension [length]$^{-3}$, the reciprocal of a volume in the $\vec{h}$-space. $\Gamma_{\alpha\vec{h}}$ is thus an intensive variable in the $\vec{h}$-space and an extensive variable in the physical space. The number $N$ of strands is the sum of the numbers $N_\alpha$ of strands in connection state $\alpha$

$$N = \sum_\alpha N_\alpha \quad ; \quad N_\alpha = \int d^3\vec{h}\, \Gamma_{\alpha\vec{h}} \tag{1}$$

Here and in the following, the sum $\sum_\alpha'$ runs over the states $\alpha=F$ and $\alpha=C$, and the integrals with respect to $\vec{h}$ extend over the whole $\vec{h}$ space. We now restrict our attention to systems homogeneous enough for the diffusion in the physical space to be negligible. Since in addition the solvent molecules, as well as the strands, are neither created nor destroyed, their respective numbers are constants:

$$dN_S = 0 \quad \text{and} \quad dN = dN_F + dN_C = 0 \tag{2}$$

The thermodynamic state of the sample is determined by its volume $V$, its temperature $T$, the distribution function $\Gamma_{\alpha\vec{h}}$ and the number $N_S$. Since $N_S$ is a constant, the Gibbs equation for the internal energy $E$ is written [Prigogine and Mazur (1953), de Groot and Mazur (1984)]

$$dE = TdS - PdV + \sum_{\alpha} \int d^3\vec{h} \, \mu_{\alpha\vec{h}} d\Gamma_{\alpha\vec{h}} \qquad (3)$$

where $\mu_{\alpha\vec{h}}$ is the chemical potential associated to an $(\alpha\vec{h})$-strand, $S$ is the entropy, $P$ is the pressure and $T$ is the absolute temperature. $\mu_{\alpha\vec{h}}$ reduces to the usual chemical potential per molecule whenever strands identify with molecules. In any case, $\mu_{\alpha\vec{h}}$ is the quantity thermodynamically conjugate to $\Gamma_{\alpha\vec{h}}$. Eq.(3) is written for independent variations $dV$ and $d\Gamma_{\alpha\vec{h}}$, this means that two kinds of external forces are separately acting on the sample : first, the pressure, conjugate to the volume, and second, the forces acting on the network, changing the connector $\vec{h}$ and the distribution $\Gamma_{C\vec{h}}$ of connected strands. Note that the volume is a thermodynamic variable but the shape is not [Landau and Lifshitz (1980)], in that a transformation which changes the shape but brings $V$, $T$ and $\Gamma_{a\vec{h}}$ back to their original values will leave all thermodynamic functions unchanged.

The macroscopic deformation must now be considered. Let $\vec{x}$ be the position vector of a given material point at time $t$, and let $\vec{x}'$ be its value at time $t'$. The deformation gradient is the tensor relating the differentials $d\vec{x}$ and $d\vec{x}'$:

$$F(t,t')_{ij} \equiv \frac{\partial x_i}{\partial x'_j} \quad ; \quad \overline{\overline{F}}(t,t') = \overline{\overline{F}}(t',t)^{-1} \qquad (4)$$

Throughout this paper, Cartesian indices will be denoted $i$, $j$, $k$ and $l$, and will be submitted to the summation convention. The velocity gradient $\overline{\overline{\kappa}}$ is related to $\overline{\overline{F}}(t,t')$ considered as a function of the present time $t$, the reference time $t'$ being kept constant, according to

$$\kappa_{ij} \equiv \frac{\partial v_i}{\partial x_j} = \frac{\partial F(t,t')_{ik}}{\partial t} F(t,t')^{-1}_{kj} \qquad (5)$$

Because of the absence of diffusion in the physical space, the network must, on average, follow the macroscopic deformation. The affine deformation assumption we now introduce states that connected strands deform affinely to the macroscopic deformation, their connector $\vec{h}$ at time $t$ being linked to their value $\vec{h}'$ at time $t'$ by the same relations as the differentials $d\vec{x}$ and $d\vec{x}'$

$$h_i = F_{ij}(t,t')h'_j \quad \text{and} \quad \frac{dh_i}{dt} = \frac{\partial F_{ij}(t,t')}{\partial t} h'_j = \kappa_{ij} h_j \qquad (6)$$

provided that the strand stays connected during time interval $(t, t')$. The affine deformation assumption amounts to linking the two sets of forces previously considered by a coherency constraint in the sense of Sekimoto (1991), ensuring that the solvent does not flow through the network. In short, we consider the limit of strong friction between network and solvent.

Macroscopic deformations thus change the distribution of connected strands. In addition, the distribution of both free and connected strands varies under the action of the kinetic relaxation mechanism to be introduced later in section 4, namely the Brownian motion and the connection/disconnection reactions. The variation $d\Gamma_{\alpha\vec{h}}$ in an infinitesimal deformation will therefore be split into two terms

$$d\Gamma_{F\vec{h}} = d^R\Gamma_{F\vec{h}}$$
$$d\Gamma_{C\vec{h}} = d^A\Gamma_{C\vec{h}} + d^R\Gamma_{C\vec{h}}$$
(7)

where $d^R\Gamma_{\alpha\vec{h}}$ accounts for the still unspecified relaxation mechanisms. The affine variation of the connector $\vec{h}$ amounts to a flow in the $\vec{h}$-space, so $d^A\Gamma_{\alpha\vec{h}}$ will be written

$$\frac{d^A\Gamma_{C\vec{h}}}{dt} = -\frac{\partial}{\partial h_i}\left(\Gamma_{C\vec{h}}h_j\kappa_{ij}\right) = -\frac{\partial\Gamma_{C\vec{h}}}{\partial h_i}h_j\kappa_{ij} - \Gamma_{C\vec{h}}\kappa_{kk}$$
(8)

The first term of the right member stems from the affine variation of the connector, and the second term accounts for the variation of the volume element $d^3\vec{h}$.

The extra-stress tensor $\bar{\bar{\tau}}$ is defined as

$$\tau_{ij} = -\frac{1}{V}\int d^3\vec{h}\,\mu_{C\vec{h}}\left(\frac{\partial\Gamma_{C\vec{h}}}{\partial h_i}h_j + \delta_{ij}\Gamma_{C\vec{h}}\right)$$
(9)

and, assuming that $\Gamma_{C\vec{h}}$ vanishes fast enough as $\vec{h}$ goes to infinity, an integration by parts yields

$$\tau_{ij} = \frac{1}{V}\int d^3\vec{h}\,\Gamma_{C\vec{h}}\frac{\partial\mu_{C\vec{h}}}{\partial h_i}h_j$$
(10)

The contribution of $d^A\Gamma_{C\vec{h}}$ to the Gibbs equation (3) can then be written as

$$\int d^3\vec{h}\,\mu_{C\vec{h}}d^A\Gamma_{C\vec{h}} = V\tau_{ij}\kappa_{ij}\,dt$$
(11)

and the Gibbs equation then becomes

$$dE = TdS + V\sigma_{ij}\kappa_{ij}dt + \sum_\alpha \int d^3\vec{h}\,\mu_{\alpha\vec{h}}\,d^R\Gamma_{\alpha\vec{h}}$$
(12)

for an infinitesimal change taking place within time interval $dt$, where the total stress tensor

$$\sigma_{ij} = -P\delta_{ij} + \tau_{ij}$$
(13)

is the quantity conjugate to $V\kappa_{ij}dt$. The quantity $-P\delta_{ij}V\kappa_{ij}dt = -PdV$ is the work performed by the pressure, and since $V\tau_{ij}\kappa_{ij}dt$ is the work performed in affinely deforming the $\vec{h}$ vectors, the extra stress $\bar{\bar{\tau}}$ is the stress associated with the variable $\vec{h}$. The stress $\bar{\bar{\sigma}}$ thus represents the total effect of the mechanical forces conjugate to both the network deformation and the macroscopic deformation of the sample, locked with one another by Sekimoto's coherency constraint resulting in the affine deformation assumption.

Expression (13) is derived under very general hypotheses that do not imply that $\bar{\bar{\tau}}$ is symmetric. A non-symmetric $\bar{\bar{\tau}}$ can result, for instance, from the torque an electric field imposes on strands having an anisotropic polarisability, contributing an orientation-dependent term to their chemical potential. A sufficient condition for the symmetry of $\bar{\bar{\tau}}$ will be given in the seventh section.

Introducing the heat flux density $\vec{q}$ allows the first principle to be written as

$$\frac{dE}{dt} = -V \operatorname{div} \vec{q} + V\sigma_{ij}\kappa_{ij} \tag{14}$$

where the quantity $V\,dt\,\operatorname{div}\vec{q}$ is the heat flowing out of volume $V$ during time interval $dt$. Volume $V$ is taken small enough for $\operatorname{div}\vec{q}$ to be practically homogeneous within $V$. This also holds for $\operatorname{div}(\vec{q}/T)$ and $\vec{q}\cdot\vec{\nabla}T^{-1}$, so using eq.(12) and (14) permits to split $dS$ into a flux term and a source term accounting for irreversible processes:

$$\frac{dS}{dt} = -V \operatorname{div} \frac{\vec{q}}{T} + \frac{d^{irr}S}{dt} \tag{15}$$

where the entropy source reads

$$\frac{d^{irr}S}{dt} = V \vec{q} \cdot \vec{\nabla}\frac{1}{T} - \frac{1}{T}\sum_{\alpha}\int d^3\vec{h}\,\mu_{\alpha\vec{h}}\frac{d^R\Gamma_{\alpha\vec{h}}}{dt} \tag{16}$$

The term $\operatorname{div}(\vec{q}/T)$ is the divergence of the entropy conduction flux, and $\vec{q}\cdot\vec{\nabla}T^{-1}$ is the rate of entropy production per unit volume due to heat conduction [de Groot and Mazur (1984)].

The irreversible phenomena, namely the heat flux $\vec{q}$, the Brownian diffusion flux of free strands in the $\vec{h}$ space and the connection/disconnection reaction rate, obey a set of constitutive equations which is left undefined up to the fourth section of this paper, where a specific model will be formulated. However, in order to comply with the second principle, the constitutive equations must satisfy the Clausius-Duhem inequality

$$\frac{d^{irr}S}{dt} \geq 0 \tag{17}$$

ensuring that the entropy source is non-negative, whatever the thermomechanical history. A stronger requirement, valid in the case of Brownian diffusion [Prigogine and Mazur (1953), de Groot and Mazur (1984)], states that the integrated term of eq.(16) must be non-negative rather than the integral as a whole, so the entropy generated within the volume element $d^3\vec{h}$ is non-negative. This condition is satisfied by the kinetic equations presented in the fourth section of this paper.

A direct consequence of inequality (17) is that the isothermal steady flow viscosity $\eta(\bar{\bar{\kappa}}) = \sigma_{ij}\kappa_{ij}/\kappa_{ij}\kappa_{ij}$ must be non negative, provided that the system is able to reach a stationary state characterised by steady values of the thermodynamic variables. The inequality $\eta(\bar{\bar{\kappa}}) \geq 0$ follows from the stationarity conditions $dE/dt = 0$ and $dS/dt = 0$, inserted into eq.(15) and (16) respectively. The volume is a thermodynamic variable, therefore it must altered by the flow, hence the condition $\kappa_{ll} = 0$.

The thermodynamic completeness of our formalism permits to derive a heat equation. The heat released by the sample can be obtained from eq.(15), by inserting the differential $dS$ evaluated as a function of $T$, $V$, and $\Gamma_{\alpha\vec{h}}$:

$$dS(T,V,\{\Gamma_{\alpha\vec{h}}\}) = \frac{V}{T}C_{V\{\Gamma\}}dT + \left(\frac{\partial P}{\partial T}\right)_{V,\{\Gamma\}}dV - \sum_\alpha \int d^3\vec{h}\left(\frac{\partial \mu_{\alpha\vec{h}}}{\partial T}\right)_{V,\{\Gamma\}}d\Gamma_{\alpha\vec{h}} \quad (18)$$

where the subscript $\{\Gamma\}$ means that the partial derivative is performed at constant distribution $\Gamma_{\alpha\vec{h}}$. The first term of the right member involves the heat capacity per unit volume, at constant volume and distribution,

$$C_{V\{\Gamma\}} = \frac{T}{V}\left(\frac{\partial S}{\partial T}\right)_{V,\{\Gamma\}} \quad (19)$$

The second term follows from a Maxwell relation involving the free energy $A(T,V,\{\Gamma_{\alpha\vec{h}}\}) \equiv E - TS$:

$$\left(\frac{\partial S}{\partial V}\right)_{T,\{\Gamma\}} = -\left(\frac{\partial^2 A}{\partial V \partial T}\right)_{\{\Gamma\}} = \left(\frac{\partial P}{\partial T}\right)_{V,\{\Gamma\}} \quad (20)$$

and another Maxwell relation allows writing the integrand of the third term as

$$\left(\frac{\delta S}{\delta \Gamma_{\alpha\vec{h}}}\right)_{T,V} = -\left(\frac{\delta \partial A}{\delta \Gamma_{\alpha\vec{h}} \partial T}\right)_V = -\left(\frac{\partial \mu_{\alpha\vec{h}}}{\partial T}\right)_{V,\{\Gamma\}} \quad (21)$$

where $\delta S/\delta \Gamma_{\alpha\vec{h}}$ denotes the functional derivative of $S$ with respect to $\Gamma_{\alpha\vec{h}}$ [Courant and Hilbert (1953)]. Finally, splitting $d\Gamma_{\alpha\vec{h}}$ according to eq.(7)-(8) yields

$$dS(T,V,\{\Gamma_{\alpha\vec{h}}\}) = \frac{V}{T}C_{V\{\Gamma\}}dT - \frac{V}{T}s_{\{\Gamma\}ij}\kappa_{ij}dt - \sum_\alpha \int d^3\vec{h}\left(\frac{\partial \mu_{\alpha\vec{h}}}{\partial T}\right)_{V,\{\Gamma\}}d^R\Gamma_{\alpha\vec{h}} \quad (22)$$

where we introduce $\bar{\bar{s}}_{\{\Gamma\}}$, the entropic stress tensor at constant distribution, such that

$$s_{\{\Gamma\}ij}\kappa_{ij} = \frac{T}{V}\int d^3\vec{h}\left(\frac{\partial \mu_{C\vec{h}}}{\partial T}\right)_{V,\{\Gamma\}}\frac{d^A\Gamma_{\vec{h}}}{dt} - \frac{T}{V}\left(\frac{\partial P}{\partial T}\right)_{V,\{\Gamma\}}\frac{dV}{dt} \quad (23)$$

$\bar{\bar{s}}_{\{\Gamma\}}$ is then the logarithmic derivative of the stress with respect to the temperature at constant volume and distribution :

$$s_{\{\Gamma\}ij} \equiv T\left(\frac{\partial \sigma_{ij}}{\partial T}\right)_{V,\{\Gamma\}} = -T\left(\frac{\partial P}{\partial T}\right)_{V,\{\Gamma\}}\delta_{ij} + t_{ij} \quad (24)$$

where the tensor

$$t_{ij} \equiv T\left(\frac{\partial \tau_{ij}}{\partial T}\right)_{V,\{\Gamma\}} = \frac{T}{V}\int d^3\vec{h}\,\Gamma_{C\vec{h}}h_j\frac{\partial}{\partial h_i}\left(\frac{\partial \mu_{C\vec{h}}}{\partial T}\right)_{V,\{\Gamma\}} \quad (25)$$

is accordingly named the entropic extra stress tensor. It will be shown in section 5 of this paper that systems that exhibit entropic strand elasticity are such that $\bar{\bar{t}} = \bar{\bar{\tau}}$, i.e. their extra stress is entirely entropic.

The rate of heat release can be written, from the expression (15) and (22) of *dS*, as

$$\operatorname{div}\vec{q} = -C_{V\{\Gamma\}}\frac{dT}{dt} + s_{\{\Gamma\}ij}\kappa_{ij} + \frac{1}{V}\sum_{\alpha}\int d^3\vec{h}\left[T\left(\frac{\partial \mu_{\alpha\vec{h}}}{\partial T}\right)_{N,\{\Gamma\}} - \mu_{\alpha\vec{h}}\right]\frac{d^R\Gamma_{\alpha\vec{h}}}{dt} \quad (26)$$

If the deformation rate is faster than the relaxation mechanisms, then the factor $d^R\Gamma_{\alpha\vec{h}}/dt$ can be neglected in eq.(22) and (26), which then reduce to $\operatorname{div}\vec{q} = -T/V\, dS/dt = -C_{V\{\Gamma\}}dT/dt + s_{\{\Gamma\}ij}\kappa_{ij}$. The entropic stress thus governs the transformation of mechanical energy into heat in fast deformations. However, it may happen that some of the relaxation mechanisms are fast, at least faster than the deformation rate, like the Brownian diffusion of free strands in the model presented in section 4. The preceding interpretation will then be shown to hold with modified coefficients of $dT/dt$ and $\overline{\overline{\kappa}}$.

**2/ Ideal systems**

The chemical potential of strands must now be explicited as a function of the thermodynamic variables. The simplest case is that of non-interacting strands constituting an ideal system, where the chemical potential takes the form [Prigogine and Mazur (1953), de Groot and Mazur (1984) § X-6]

$$\mu_{\alpha\vec{h}}(P,T,\Gamma_{\alpha\vec{h}}) = \zeta_{\alpha\vec{h}}(P,T) + k_B T \ln\frac{\theta\,\Gamma_{\alpha\vec{h}}}{N} \tag{27}$$

where $k_B$ is the Boltzmann constant, and the constant $\theta$ is a volume of the $\vec{h}$-space, making the argument of the logarithm non-dimensional. The standard chemical potential $\zeta_{\alpha\vec{h}}$ is the value the chemical potential takes when the concentration $\Gamma_{\alpha\vec{h}}/N$ amounts to one strand per volume $\theta$ of the $\vec{h}$-space. $\zeta_{\alpha\vec{h}}$ thus accounts for the interactions of the strand with its surrounding molecules, whilst $k_B T \ln\theta\,\Gamma_{\alpha\vec{h}}/N$ is the combinatoric entropy of the strand distribution. In the following, $\zeta_{\alpha\vec{h}}$ will be called simply the connector potential. Since the system is ideal, $\zeta_{\alpha\vec{h}}$ depends on the intensive variables $T$ and $P$ (note that at constant $T$ and $P$, the volume varies with the strand distribution, the strand volume $V_{\alpha\vec{h}} = \left(\partial\mu_{\alpha\vec{h}}/\partial P\right)_{T,\{\Gamma\}}$ depending on $\alpha$ and $\vec{h}$. $\zeta_{\alpha\vec{h}}$ is thus a function of $T$, $V$ and of the whole distribution $\{\Gamma\}$).

The gradient of the chemical potential in the $\vec{h}$-space splits into two forces [Jongschaap & al., (1997)]

$$\frac{\partial\mu_{\alpha\vec{h}}}{\partial\vec{h}} = \vec{f}_{\alpha\vec{h}} + k_B T \frac{\partial\ln\Gamma_{\alpha\vec{h}}}{\partial\vec{h}} \tag{28}$$

The first term is the connector force

$$\vec{f}_{\alpha\vec{h}} = \frac{\partial\zeta_{\alpha\vec{h}}}{\partial\vec{h}} \tag{29}$$

deriving from the connector potential, and defined with the customary positive sign. The second term is the so-called the Brownian force, coming from the combinatoric entropy. Inserting expression (28) into the definition (10) of the extra stress tensor yields the familiar expression

$$\tau_{ij} = \frac{1}{V}\int d^3\vec{h}\,\Gamma_{C\vec{h}}\left(f_{C\vec{h}i}h_j - k_B T \delta_{ij}\right) \tag{30}$$

The entropic extra stress $\bar{\bar{t}}$, defined in eq.(25), has the same form as $\bar{\bar{\tau}}$ in eq.(10) with $\partial\mu_{\alpha\vec{h}}/\partial\vec{h}$ substituted for

$$T\frac{\partial}{\partial\vec{h}}\left(\frac{\partial\mu_{\alpha\vec{h}}}{\partial T}\right)_{V,\{\Gamma\}} = T\left(\frac{\partial\vec{f}_{\alpha\vec{h}}}{\partial T}\right)_V + k_B T\frac{\partial\ln\Gamma_{\alpha\vec{h}}}{\partial\vec{h}} \tag{31}$$

The first term is the entropic part of the connector force and the second term is just the Brownian force, which is entirely entropic. The entropic extra stress then reads

$$t_{ij} = \frac{T}{V}\int d^3\vec{h}\,\Gamma_{C\vec{h}}\left(\left(\frac{\partial f_{C\vec{h}i}}{\partial T}\right)_V h_j - k_B \delta_{ij}\right) \tag{32}$$

**3/ Equilibrium**

Since Brownian diffusion and connection/disconnections allow the strands to reach every possible state $(\alpha, \vec{h})$, the chemical potential at equilibrium cannot depend on $\alpha$ nor $\vec{h}$:

$$\mu_{eq}(P,T) = \mu_{\alpha\vec{h}}^{eq} = \zeta_{\alpha\vec{h}}(P,T) + k_B T \ln \frac{\theta \Gamma_{\alpha\vec{h}}^{eq}}{N} \tag{33}$$

Inserting this relation into (10) gives the equilibrium extra stress tensor:

$$\bar{\bar{\tau}}^{eq} = 0 \tag{34}$$

There is no such general relation for the entropic extra stress at equilibrium $\bar{\bar{t}}^{eq}$, because equation (33), stating that $\mu_{\alpha\vec{h}}$ is uniform in the $(\alpha, \vec{h})$-space, does not imply that $\left(\partial \mu_{\alpha\vec{h}} / \partial T\right)_{V,\{\Gamma\}}$ also is uniform. An example of this situation is given about eq.(87).

The equilibrium distribution function obeys the Boltzmann statistics

$$\Gamma_{\alpha\vec{h}}^{eq} = \frac{N}{\theta} \exp\left(\frac{\mu_{eq} - \zeta_{\alpha\vec{h}}}{k_B T}\right) \tag{35}$$

We define for further use the normalised equilibrium distribution $\phi_{\alpha\vec{h}}$, such that $\int d^3\vec{h} \phi_{\alpha\vec{h}} = 1$

$$\phi_{\alpha\vec{h}} \equiv \frac{\Gamma_{\alpha\vec{h}}^{eq}}{N_{\alpha}^{eq}} = \frac{\exp\frac{-\zeta_{\alpha\vec{h}}}{k_B T}}{\int d^3\vec{h}' \exp\frac{-\zeta_{\alpha\vec{h}'}}{k_B T}} \tag{36}$$

It is linked to the connector force by the relation

$$\vec{f}_{\alpha\vec{h}} = -k_B T \frac{\partial \ln \phi_{\alpha\vec{h}}}{\partial \vec{h}} \tag{37}$$

Introducing the connection potential $\Delta\zeta_{\vec{h}} \equiv \zeta_{C\vec{h}} - \zeta_{F\vec{h}}$ as the difference of connector potentials with the same connector, and using the equilibrium condition (33), we rewrite the ratio $\Gamma_{F\vec{h}}^{eq} / \Gamma_{C\vec{h}}^{eq}$ of equilibrium distributions according to the mass action law

$$\frac{\Gamma_{F\vec{h}}^{eq}}{\Gamma_{C\vec{h}}^{eq}} = \exp\left(\frac{\Delta\zeta_{\vec{h}}}{k_B T}\right) \tag{38}$$

In the important case of an $\vec{h}$-independent connection potential, the equilibrium distribution of connected strands and that of the free strands have the same dependence on $\vec{h}$.

**4/ Kinetics**

The formulation of kinetic equations for the dissipative processes $d^R\Gamma_{\alpha\vec{h}}$ and $\vec{q}$ permits to derive explicit rheothermodynamic constitutive equations, giving the thermodynamic functions, the distribution function, the stress and the heat production as functions of the deformation and temperature history. These kinetic equations must result in a non-negative source of entropy, in order to comply with the second principle.

We first formulate the kinetic equations. The connection and the disconnection of strands are assumed to take place with no change of $\vec{h}$, fast enough to be considered instantaneous at the thermorheological timescale. This allows treating the connections and the disconnections as chemical reactions, in the sense that a chemical reaction implies an abrupt change of the objects that react. Choosing the connection as the direct reaction and the disconnection as the reverse reaction, we write symbolically

$$(F,\vec{h}) \underset{}{\overset{v_{\vec{h}}}{\rightleftarrows}} (C,\vec{h}) \qquad (39)$$

where the reaction rate $v_{\vec{h}}$ numbers the free strands which connect minus the connected strands which disconnect, per unit time and per unit volume in the $\vec{h}$ space. The rate $v_{\vec{h}}$ is assumed to obey a first order kinetics in the chemical sense [de Groot and Mazur (1984)], according to

$$v_{\vec{h}} = \beta_{C\vec{h}}\Gamma_{F\vec{h}} - \beta_{F\vec{h}}\Gamma_{C\vec{h}} \qquad (40)$$

where $\beta_{C\vec{h}}$ and $\beta_{F\vec{h}}$ are the connection and disconnection rate constants, respectively, with the dimension of an inverse time. Since our system is assumed ideal, the rate constants depend on volume and temperature but not on the strand distribution, the strand being non-interacting. This excludes variation of the rate constants with the stress or with a structural parameter.

In addition to the connection/disconnection reactions, free strands undergo a Brownian diffusion in the $\vec{h}$ space, described by the diffusion flux $\vec{w}_{\vec{h}}$. Its $i$-component $w_{\vec{h}i}$ is the number of free strands having their connector diffusing in the $i$-direction of the $\vec{h}$ space, per unit area across this $i$-direction and per unit time. This Brownian diffusion is assumed to be so fast as to establish a partial equilibrium with respect to $\vec{h}$, with the consequence that the chemical potential of the free strands does not depend on $\vec{h}$,

$$\mu_{F\vec{h}} = \mu_F \qquad (41)$$

though $\mu_F$ can differ from the equilibrium value $\mu_{eq}$.

The variation rate of the distribution function due to the relaxation processes can now be written

$$\frac{d^R \Gamma_{F\vec{h}}}{dt} = -v_{\vec{h}} - \frac{\partial}{\partial h_i} w_{\vec{h}i}$$

$$\frac{d^R \Gamma_{C\vec{h}}}{dt} = v_{\vec{h}}$$

(42)

At equilibrium, the Brownian flux $\vec{w}_{\vec{h}}^{eq}$ and the connection rate $v_{\vec{h}}^{eq}$ satisfy the conditions

$$v_{\vec{h}}^{eq} = 0$$

$$\vec{w}_{\vec{h}}^{eq} = 0$$

(43)

These separate equalities, as well as the condition that $\vec{w}_{\vec{h}}^{eq}$ vanishes (instead of the weaker conditions $\partial \vec{w}_{\vec{h}}^{eq}/\partial \vec{h} \equiv 0$) follow from the principle of detailed balance [de Groot and Mazur (1984)]. The first line of eq.(43) implies that the ratio $\beta_{F\vec{h}}/\beta_{C\vec{h}}$ cannot be chosen at will, it must satisfy the detailed balance condition

$$\beta_{C\vec{h}} \Gamma_{F\vec{h}}^{eq} = \beta_{F\vec{h}} \Gamma_{C\vec{h}}^{eq}$$

(44)

The mass action law (38) allows formulating the detailed balance in terms of the connection potential:

$$\frac{\beta_{F\vec{h}}}{\beta_{C\vec{h}}} = \frac{\Gamma_{F\vec{h}}^{eq}}{\Gamma_{C\vec{h}}^{eq}} = \exp\left(\frac{\Delta \zeta_{\vec{h}}}{k_B T}\right)$$

(45)

One more constitutive equation must be introduced, that for the heat flux $\vec{q}$. The simplest heat conduction equation, Fourier's law, involves a non-negative thermal conductivity $\lambda$:

$$\vec{q} = -\lambda \vec{\nabla} T$$

(46)

Inserting this expression into the entropy production (16) ensures that the first term of the right member is non-negative. The second term can be rewritten, using eq.(40) to (42):

$$\sum_\alpha \int d^3\vec{h} \mu_{\alpha\vec{h}} \frac{d^R \Gamma_{\alpha\vec{h}}}{dt} = \int d^3\vec{h} \left[ w_{\vec{h}i} \frac{\partial \mu_{F\vec{h}}}{\partial h_i} + v_{\vec{h}} \Delta \mu_{\vec{h}} \right]$$

$$= \int d^3\vec{h} \left( \beta_{C\vec{h}} \Gamma_{F\vec{h}} - \beta_{F\vec{h}} \Gamma_{C\vec{h}} \right) \Delta \mu_{\vec{h}}$$

(47)

where $\Delta\mu_{\vec{h}} \equiv \mu_{C\vec{h}} - \mu_{F\vec{h}} = \Delta\zeta_{\vec{h}} + k_B T \ln \Gamma_{C\vec{h}}/\Gamma_{F\vec{h}}$ is the chemical affinity associated to the connection reaction (39), and $\partial \mu_{F\vec{h}}/\partial \vec{h}$ is the affinity of the Brownian diffusion of free strands, conjugate to the diffusion flux $\vec{w}_{\vec{h}}$ [Prigogine and Mazur (1953), de Groot and Mazur (1984) § X-6]. The Brownian diffusion of free strands gives no contribution to the source of entropy because its affinity vanishes, according to eq.(41). Using relation (27) and the detailed balance (45), the connection affinity can be written

$$\Delta\mu_{\vec{h}} = k_B T \ln \frac{\beta_{F\vec{h}} \Gamma_{C\vec{h}}}{\beta_{C\vec{h}} \Gamma_{F\vec{h}}} \tag{48}$$

and inserting eq.(46)-(48) into the source of entropy (16) then yields

$$\frac{d^{irr}S}{dt} = \lambda \frac{V}{T^2}(\vec{\nabla}T)^2 + k_B \int d^3\vec{h} \left(\beta_{C\vec{h}} \Gamma_{F\vec{h}} - \beta_{F\vec{h}} \Gamma_{C\vec{h}}\right) \ln \frac{\beta_{C\vec{h}} \Gamma_{F\vec{h}}}{\beta_{F\vec{h}} \Gamma_{C\vec{h}}} \tag{49}$$

Obviously, since the logarithm is an increasing function of its argument, the integrand is non-negative. Our model therefore obeys the Clausius-Duhem form (17) of the second principle :

$$\frac{d^{irr}S}{dt} \geq 0 \tag{50}$$

stating that the entropy production is non-negative. This production is zero at equilibrium, when the temperature gradient vanishes as well as the integrated term of eq.(49), as a consequence of the detailed balance condition (44). Furthermore, the fact that the integrated term itself is non-negative ensures that any volume element $d^3\vec{h}$ generates a non-negative entropy [Prigogine and Mazur (1953)].

The distribution functions must now be related to the thermomechanical history. Using eq. (41) and (27), the distribution of free strands can be written as the product of the total number $N_F$ of free strands times the normalised distribution $\phi_{F\vec{h}}$ defined in (36) :

$$\Gamma_{F\vec{h}} = \frac{N}{\theta} \exp\left(\frac{\mu_F - \zeta_{F\vec{h}}}{k_B T}\right) = N_F \phi_{F\vec{h}} \tag{51}$$

The derivation of the distribution of connected strands is more involved. According to the affine deformation assumption, a connected strand of connector $\vec{h}$ at time $t$ must have been connected to the network at some previous time $t'$, with its connector $\vec{h}'$ related to its present value $\vec{h}$ by the affine transformation (6) : $\vec{h}' = \overline{\overline{F}}(t',t) \cdot \vec{h}$. The probability that a strand connected at time $t'$ stays connected up to time $t$ reads

$$p(t|t',\vec{h}') = \exp\left(-\int_{t'}^{t} dt'' \beta_{F\vec{h}''}(t'')\right) \quad \text{with} \quad \begin{aligned} \vec{h}'' &= \overline{\overline{F}}(t'',t) \cdot \vec{h} \\ \vec{h}' &= \overline{\overline{F}}(t',t) \cdot \vec{h} \end{aligned} \tag{52}$$

where $\beta_{F\vec{h}''}(t'')$ denotes the value the disconnection rate constant had at time $t''$ (as a function of the temperature and the pressure at that time). The number of free strands which connect within volume element $d^3\vec{h}'$ during time interval $(t',t'+dt')$, and stay connected up to time $t$, can be written $\beta'_{C\vec{h}'} dt' \Gamma'_{F\vec{h}'} d^3\vec{h}' p(t|t',\vec{h}')$, where $\beta'_{C\vec{h}'}$ and $\Gamma'_{F\vec{h}'}$ are evaluated at time $t'$. Due to the affine character of the deformation, the volume element $d^3\vec{h}'$ at time $t'$ becomes $d^3\vec{h}$ at time $t$, such that $d^3\vec{h}' = d^3\vec{h} \det \overline{\overline{F}}$. Integrating along the past history of a given volume element $d^3\vec{h}$ then yields the distribution of connected

strands at time *t* :

$$\Gamma_{C\vec{h}} = \int_{-\infty}^{t} dt' \det\overline{\overline{F}}\ N'_F \phi'_{F\vec{h}'} \beta'_{C\vec{h}'} p(t|t',\vec{h}')\quad (53)$$

where $\vec{h}' = \overline{\overline{F}}\cdot\vec{h}$. Hereafter, all primed quantities are evaluated at the past time *t'*. Since temperature and pressure vary throughout the deformation history, the temperature and pressure-dependent quantities $\phi'_{F\vec{h}'}$ and $\beta'_{C\vec{h}'}$ will vary accordingly.

Integrating eq.(53) over $\vec{h}'$ yields an integral equation for the numbers $N_F$ and $N_C$ :

$$N_C = N - N_F = \int_{-\infty}^{t} dt' N'_F \int d^3\vec{h}'\phi'_{F\vec{h}'}\beta'_{C\vec{h}'}p(t|t',\vec{h}')\quad (54)$$

Differentiating with respect to *t*, or equivalently integrating eq.(42) over $\vec{h}$, gives the total connection rate:

$$\frac{dN_C}{dt} = -\frac{dN_F}{dt} = \int d^3\vec{h}\, v_{\vec{h}} = N_F \int d^3\vec{h}\, \beta_{C\vec{h}}\phi_{F\vec{h}} - \int d^3\vec{h}\, \beta_{F\vec{h}}\Gamma_{C\vec{h}}\quad (55)$$

Equations (51) and (53)-(55) permit to derive the distribution $\Gamma_{\alpha\vec{h}}$ from the thermomechanical history, thus determining the complete set of thermodynamic functions. We now focus on the stress and the heat production. Inserting the distribution (53) of connected strands into expression (30) and (32) and integrating over $d^3\vec{h}' = \det\overline{\overline{F}}\, d^3\vec{h}$ yields

$$\tau_{ij} = \frac{1}{V}\int_{-\infty}^{t} dt' \int d^3\vec{h}'\left\{\left(f_{C\vec{h}i}h_j - k_B T\delta_{ij}\right)N'_F\phi'_{F\vec{h}'}\beta'_{C\vec{h}'}p(t|t',\vec{h}')\right\}$$

$$t_{ij} = \frac{T}{V}\int_{-\infty}^{t} dt' \int d^3\vec{h}'\left\{\left(\left(\frac{\partial f_{C\vec{h}i}}{\partial T}\right)_V h_j - k_B\delta_{ij}\right)N'_F\phi'_{F\vec{h}'}\beta'_{C\vec{h}'}p(t|t',\vec{h}')\right\}\quad (56)$$

The force $\vec{f}_{C\vec{h}} = \partial\zeta_{C\vec{h}}/\partial\vec{h}$ and the temperature *T* are of course evaluated at time *t*.

Consider now the heat production. Because of its instantaneous character, the contribution of the Brownian diffusion of free strands to eq.(26) can be lumped into the coefficients of $dT/dt$ and $\overline{\overline{\kappa}}$: using eq.(51) and accounting for the volume and temperature dependence of $\phi_{F\vec{h}}$, the time derivative $d\Gamma_{F\vec{h}}/dt$ can be written

$$\frac{d\Gamma_{F\vec{h}}}{dt} = \phi_{F\vec{h}}\frac{dN_F}{dt} + N_F\left(\frac{\partial\phi_{F\vec{h}}}{\partial V}\right)_T V\kappa_{kk} + N_F\left(\frac{\partial\phi_{F\vec{h}}}{\partial T}\right)_V \frac{dT}{dt}\quad (57)$$

and the variation of entropy (18)-(22) becomes

$$\frac{dS}{dt} = \frac{V}{T}C_V\frac{dT}{dt} - \frac{V}{T}s_{ij}\kappa_{ij} - \int d^3\vec{h}\left[\frac{dN_F}{dt}\phi_{F\vec{h}}\left(\frac{\partial\mu_{F\vec{h}}}{\partial T}\right)_{V,\{\Gamma\}} + v_{\vec{h}}\left(\frac{\partial\mu_{C\vec{h}}}{\partial T}\right)_{V,\{\Gamma\}}\right]\quad (58)$$

where, from eq.(55), the integrated term is contributed by the connection/disconnection reactions. The quantity

$$C_V = C_{V\{\Gamma\}} - \frac{N_F}{V} T \int d^3\vec{h} \left(\frac{\partial \mu_{F\vec{h}}}{\partial T}\right)_{V,\{\Gamma\}} \left(\frac{\partial \phi_{F\vec{h}}}{\partial T}\right)_V \tag{59}$$

is the heat capacity at fixed distribution of the connected strands, but allowing for the fast diffusion of the free strands, and the tensor

$$s_{ij} = s_{\{\Gamma\}ij} + \delta_{ij} N_F T \int d^3\vec{h} \left(\frac{\partial \mu_{F\vec{h}}}{\partial T}\right)_{V,\{\Gamma\}} \left(\frac{\partial \phi_{F\vec{h}}}{\partial V}\right)_T$$

$$= t_{ij} + T\delta_{ij} \left[ -\left(\frac{\partial P}{\partial T}\right)_{V,\{\Gamma\}} + N_F \int d^3\vec{h} \left(\frac{\partial \mu_{F\vec{h}}}{\partial T}\right)_{V,\{\Gamma\}} \left(\frac{\partial \phi_{F\vec{h}}}{\partial V}\right)_T \right] \tag{60}$$

is the entropic stress tensor, also allowing for diffusion of the free strands, which contributes an isotropic term.

The rate of heat production is obtained by inserting (58) into eq.(15) and using (41), giving

$$\text{div}\,\vec{q} = -C_V \frac{dT}{dt} + s_{ij}\kappa_{ij} + \frac{1}{V} \int d^3\vec{h} \left( v_{\vec{h}} T \left(\frac{\partial \mu_{C\vec{h}}}{\partial T}\right)_{V,\{\Gamma\}} + \frac{dN_F}{dt} \phi_{F\vec{h}} T \left(\frac{\partial \mu_{F\vec{h}}}{\partial T}\right)_{V,\{\Gamma\}} - v_{\vec{h}} \Delta \mu_{\vec{h}} \right) \tag{61}$$

In deformations and temperature variations faster than the reaction rate constants $\beta_{\alpha\vec{h}}$, but slower than the Brownian diffusion of free strands, the integral term of eq.(61) becomes negligible, as well as that of eq.(58). Then both equations reduce to

$$\text{div}\,\vec{q} = \frac{1}{V}\frac{dE}{dt} - \sigma_{ij}\kappa_{ij}$$

$$= -\frac{T}{V}\frac{dS}{dt} = -C_V \frac{dT}{dt} + s_{ij}\kappa_{ij} \tag{62}$$

revealing the reversible thermoelastic behaviour of the network in fast deformations.

## 5/ Entropic strands : application to polymers

To a good approximation, polymeric strands can be modelled by entropic chains with a configurational entropy depending on $\vec{h}$ only, and with $\vec{h}$-independent volume and internal energy. This situation is described by the chemical potential

$$\mu_{\alpha\vec{h}} = \zeta_{\alpha\vec{h}} + k_B T \ln \frac{\theta \Gamma_{\alpha\vec{h}}}{N} \qquad (63)$$

$$\zeta_{\alpha\vec{h}} = \zeta_\alpha(P,T) - k_B T \ln \Omega_{\vec{h}}$$

where $\Omega_{\vec{h}}$, the number of accessible configurations for a given $\vec{h}$, is $(P,T)$-independent, and $\zeta_\alpha(P,T)$ does not depend on $\vec{h}$. The quantity $\Omega_{\vec{h}}$ appears in the strand entropy, given by the standard relation [de Groot and Mazur (1984)]

$$S_{\alpha\vec{h}} \equiv -\left(\frac{\partial \mu_{\alpha\vec{h}}}{\partial T}\right)_{P,\{\Gamma\}} = -\left(\frac{\partial \zeta_a}{\partial T}\right)_P + k_B \ln \Omega_{\vec{h}} - k_B \ln \frac{\theta \Gamma_{\alpha\vec{h}}}{N} \qquad (64)$$

$k_B \ln \Omega_{\vec{h}}$ being the configurational entropy. It is easy to check that expression (63) effectively leads to $\vec{h}$-independent strand volume $V_\alpha = \left(\partial \mu_{\alpha\vec{h}}/\partial P\right)_{T\{\Gamma\}} = \left(\partial \zeta_{\alpha\vec{h}}/\partial P\right)_T$ and strand energy $E_\alpha = \mu_{\alpha\vec{h}} - PV_\alpha + TS_{\alpha\vec{h}}$. The equilibrium distribution (35) is proportional to $\Omega_{\vec{h}}$

$$\Gamma^{eq}_{\alpha\vec{h}} = N^{eq}_\alpha \phi_{\vec{h}} \;, \quad \text{with} \quad \phi_{\vec{h}} = \frac{\Omega_{\vec{h}}}{\int d^3\vec{h}'\, \Omega_{\vec{h}'}} \quad \text{and} \quad N^{eq}_\alpha = \frac{N \exp\frac{-\zeta_\alpha}{k_B T}}{\exp\frac{-\zeta_C}{k_B T} + \exp\frac{-\zeta_F}{k_B T}} \qquad (65)$$

where $\phi_{\vec{h}}$ is $(P,T)$-independent. The detailed balance (45) then reads

$$\frac{\beta_{F\vec{h}}}{\beta_{C\vec{h}}} = \exp\frac{\Delta\zeta}{k_B T} \;; \quad \text{with} \quad \Delta\zeta = \zeta_C - \zeta_F \qquad (66)$$

and states that the ratio $\beta_{F\vec{h}}/\beta_{C\vec{h}}$ is $\vec{h}$-independent.

The connector force derived from expression (63) is $\alpha$-independent; it derives from the configurational entropy:

$$\vec{f}_{\alpha\vec{h}} = \vec{f}_{\vec{h}} = -k_B T \frac{\partial \ln \Omega_{\vec{h}}}{\partial \vec{h}} \qquad (67)$$

It thus satisfies the relation

$$\vec{f}_{\vec{h}} = T \left(\frac{\partial \vec{f}_{\vec{h}}}{\partial T}\right)_V \qquad (68)$$

with the consequence that that the extra stress equals the entropic extra stress

$$\tau_{ij} = t_{ij} = \frac{1}{V}\int d^3\vec{h}\ \Gamma_{C\vec{h}}\left(f_i h_j - k_B T \delta_{ij}\right) \tag{69}$$

In systems of entropic strands, the free strand contribution to $\overline{\overline{t}}$ is always zero because of relations (41) and (68), and the whole entropic stress tensor vanishes at equilibrium, $\overline{\overline{t}}^{eq} = 0$.

Since the normalised distribution $\phi_{\vec{h}}$ is independent of pressure, volume and temperature, eq.(60) reduces to

$$s_{ij} = s_{\{\Gamma\}ij} = \tau_{ij} - T\left(\frac{\partial P}{\partial T}\right)_{V,N_\alpha}\delta_{ij} \tag{70}$$

i.e. the entropic stress is independent of whether the free strands are allowed to diffuse or not. A similar relation holds for the heat capacity (59):

$$C_V = C_{V\{\Gamma\}} \tag{71}$$

The following relation

$$T\left(\frac{\partial \mu_{\alpha\vec{h}}}{\partial T}\right)_{V,\{\Gamma\}} - \mu_{\alpha\vec{h}} = T\left(\frac{\partial \zeta_\alpha}{\partial T}\right)_V - \zeta_\alpha \tag{72}$$

implies that $\left(\partial \mu_{F\vec{h}}/\partial T\right)_{V,\{\Gamma\}}$ is $\vec{h}$-independent like $\mu_{F\vec{h}}$, therefore the Brownian diffusion of free strands does not contribute to eq.(22), which can be written

$$\frac{dS}{dt} = \frac{V}{T}C_V\frac{dT}{dt} - \frac{V}{T}\tau_{ij}\kappa_{ij} + V\left(\frac{\partial P}{\partial T}\right)_{V,N_\alpha}\kappa_{kk} - \int d^3\vec{h}\, v_{\vec{h}}\left(\frac{\partial \Delta\mu_{\vec{h}}}{\partial T}\right)_{V,\{\Gamma\}} \tag{73}$$

This allows rewriting the heat production equation (26) in the simple form

$$\mathrm{div}\,\vec{q} = -C_V\frac{dT}{dt} + \tau_{ij}\kappa_{ij} - T\left(\frac{\partial P}{\partial T}\right)_V \kappa_{kk} + \frac{1}{V}\left(T\left(\frac{\partial \Delta\zeta}{\partial T}\right)_V - \Delta\zeta\right)\frac{dN_C}{dt} \tag{74}$$

When considering either a permanent network or a fast deformation, the last terms of eq.(73) and (74) can be neglected, and eq.(62) results in the familiar equation of rubber thermoelasticity $\mathrm{div}\,\vec{q} = -T/V\,dS/dt = -C_V\,dT/dt + \tau_{ij}\kappa_{ij} - T(\partial P/\partial T)_V\,\kappa_{kk}$ [Sekimoto (1991)].

**6/ The limit $N_C \ll N$ : constant connection rate**

This section and the following one make use of the results of sections 1 to 4, but do not depend on section 5.

The limiting case of a number of free strands $N_F$ largely exceeding the number of connected strands $N_C$ allows some simplifications due to the fact that $N_F$, being close to the total number $N$, remains unaffected by the thermomechanical history of the sample. More precisely, one has $N_C \ll N \cong N_F \cong N_F^{eq}$ and $\Gamma_{F\vec{h}} \cong \Gamma_{F\vec{h}}^{eq}$ at lowest order in $N_C/N$. The number $\beta_{C\vec{h}}\Gamma_{F\vec{h}}$ of connection per unit time and per unit volume of the $\vec{h}$-space, thus stays equal to its equilibrium value

$$g_{\vec{h}} \equiv \beta_{C\vec{h}}\Gamma_{F\vec{h}}^{eq} \tag{75}$$

where $g_{\vec{h}}$ is called the generation function. We now rewrite the main results of section 4/ in terms of the generation function. The connection rate can be rewritten

$$\mathrm{v}_{\vec{h}} = g_{\vec{h}} - \beta_{F\vec{h}}\Gamma_{C\vec{h}} = \beta_{F\vec{h}}\left(\Gamma_{C\vec{h}}^{eq} - \Gamma_{C\vec{h}}\right) \tag{76}$$

The detailed balance (44) and the Boltzmann distribution (35) become

$$g_{\vec{h}} = \beta_{F\vec{h}}\Gamma_{C\vec{h}}^{eq} \propto \beta_{F\vec{h}} \exp\frac{-\zeta_{C\vec{h}}}{k_B T} \tag{77}$$

This relation allows rewriting the connector force as

$$\vec{f}_{C\vec{h}} = -k_B T \frac{\partial}{\partial \vec{h}} \ln \frac{g_{\vec{h}}}{\beta_{F\vec{h}}} \tag{78}$$

Relations (77)-(78) represent how the kinetic quantities $g_{\vec{h}}$ and $\beta_{F\vec{h}}$ are linked to the Boltzmann equilibrium distribution $\Gamma_{C\vec{h}}^{eq} \propto \exp-\zeta_{C\vec{h}}/k_B T$. Inserting the detailed balance (77) in the source of entropy (49) shows that the Clausius-Duhem inequality is satisfied:

$$0 \leq \frac{d^{\mathrm{irr}} S}{dt} = \lambda \frac{V}{T^2}\left(\vec{\nabla}T\right)^2 + k_B \int d^3\vec{h}\ \beta_{F\vec{h}}\left(\Gamma_{C\vec{h}}^{eq} - \Gamma_{C\vec{h}}\right)\ln\frac{\Gamma_{C\vec{h}}^{eq}}{\Gamma_{C\vec{h}}} \tag{79}$$

The expression (53) of the distribution of connected strands becomes

$$\Gamma_{C\vec{h}} = \int_{-\infty}^{t} dt'\det\overline{\overline{F}}\ g'_{\vec{h}'}\ p(t|t',\vec{h}') \tag{80}$$

where $\vec{h}' = \overline{\overline{F}}\cdot\vec{h}$, and eq.(56) reads

$$\tau_{ij} = \frac{1}{V}\int_{-\infty}^{t}dt'\int d^3\vec{h}'\left\{\left(f_{C\vec{h}\,i}h_j - k_B T\delta_{ij}\right)g'_{\vec{h}'}\ p(t|t',\vec{h}')\right\}$$

$$t_{ij} = \frac{T}{V}\int_{-\infty}^{t}dt'\int d^3\vec{h}'\left\{\left(\left(\frac{\partial f_{C\vec{h}\,i}}{\partial T}\right)_V h_j - k_B\delta_{ij}\right)g'_{\vec{h}'}\ p(t|t',\vec{h}')\right\} \tag{81}$$

## 7/ Isotropic and Gaussian systems, comparison with earlier transient network models

In this section, we give the simpler form our results take under the most usual assumptions about the connector force law and the connection/disconnection rate, and then we compare our results with the corresponding ones found in the literature.

First, consider ideal systems with isotropic $\zeta_{a\vec{h}}$, depending on the connector magnitude $h=|\vec{h}|$ only. The connector force $\vec{f}_{\alpha\vec{h}}$ and its derivative with respect to $T$ are collinear with $\vec{h}$

$$\vec{f}_{\alpha\vec{h}} = \vec{h}\, h^{-1} f_{\alpha h}\ , \quad f_{\alpha h} = \frac{\partial \zeta_{\alpha h}}{\partial h}\ ; \quad \text{and} \quad \left(\frac{\partial \vec{f}_{\alpha\vec{h}}}{\partial T}\right)_V = \vec{h}\, h^{-1} \left(\frac{\partial f_{\alpha h}}{\partial T}\right)_V \tag{82}$$

Since the strand elasticity is not assumed to be entirely entropic, relation (68) does not hold in general. Under assumption (82), the extra stress and the entropic extra stress are symmetric tensors :

$$\tau_{ij} = \tau_{ji} = \frac{1}{V}\int d^3\vec{h}\, h_i h_j \Gamma_{C\vec{h}} h^{-1} f_{Ch} - \frac{N_C}{V} k_B T \delta_{ij} \tag{83}$$

$$t_{ij} = t_{ji} = \frac{T}{V}\int d^3\vec{h}\, h_i h_j \Gamma_{C\vec{h}} h^{-1} \left(\frac{\partial f_{Ch}}{\partial T}\right)_V - \frac{N_C}{V} k_B T \delta_{ij} \tag{84}$$

Another usual assumption is that the connection/disconnection rate constant $\beta_{\alpha\vec{h}} = \beta_{\alpha h}$ is isotropic, depending on the magnitude of $\vec{h}$ only. This, added to the isotropy of $\zeta_{a\vec{h}}$, makes the system completely isotropic.

The most usual force law corresponds to Gaussian strands, which have a Hookean, $\alpha$-independent force law:

$$\zeta_{\alpha\vec{h}} = \zeta_\alpha + \tfrac{1}{2}\chi \vec{h}^2\ ; \quad \vec{f}_{\alpha\vec{h}} = \chi \vec{h}\ ; \quad \text{and} \quad \phi_{\alpha\vec{h}} = \left(\frac{\chi}{2\pi k_B T}\right)^{3/2} \exp-\frac{\chi \vec{h}^2}{2k_B T} \tag{85}$$

For instance, the spring constant is $\chi = 3k_B T/na^2$ for freely jointed entropic chains made with $n$ segments of length $a$, at moderate extensions ($h \ll na$). The connector force is the same for free and connected strands, thus the connection potential $\Delta\zeta_{\vec{h}} = \Delta\zeta$ is $\vec{h}$-independent. The extra stress (30) then takes the familiar form

$$\tau_{ij} = \frac{\chi}{V}\int d^3\vec{h}\, \Gamma_{C\vec{h}} h_i h_j - \frac{N_C}{V} k_B T \delta_{ij} \tag{86}$$

and the entropic extra stress (32) can be written

$$t_{ij} = \frac{T}{V}\frac{\partial \chi}{\partial T}\int d^3\vec{h}\, \Gamma_{C\vec{h}} h_i h_j - \frac{N_C}{V} k_B T \delta_{ij} \tag{87}$$

It is easy to see in this particular example that $\overline{\overline{t}}$ can be non-zero at equilibrium : just consider a purely energetic, hence $T$-independent spring constant $\chi$. Then the entropic extra stress reduces to a pressure, $\overline{\overline{t}} = -Nk_B T \overline{\overline{\delta}}/V$. On the other hand, the entropic elasticity of the freely jointed chain makes the spring constant proportional to $T$, with $\overline{\overline{t}}$ vanishing at equilibrium.

The transient network model presented here must now be compared with those belonging to the same lineage, namely that of Green-Tobolsky and Yamamoto. First, Green and Tobolsky (1946) considered a Gaussian generation function $g_{\vec{h}}$ with an $\vec{h}$-independent disconnection rate constant $\beta_{F\vec{h}} = \beta_F$ (using our notations). This model fits within our formalism, therefore it is thermodynamically consistent. Green and Tobolsky indeed showed that their model always leads to dissipation of mechanical work in closed-cycle deformations.

Such is not the case of Yamamoto's model (1956-1958), which allows a free choice for $\beta_{F\vec{h}}$, $g_{\vec{h}}$ and $\zeta_{C\vec{h}}$. This contradicts our equations (77)-(78) which link the connector potential, and hence the connector force, to the kinetic quantities $\beta_{F\vec{h}}$ and $g_{\vec{h}}$. Should this relation be enforced, Yamamoto's model would fit within ours. Lodge's model for Gaussian strands (1968) superposes a distribution of Green-Tobolsky networks, and shares the same consistency. In treating non-Gaussian strands, Lodge (1968) uses an inverse-Langevin connector force law, whereas he keeps a Gaussian generation function and an $\vec{h}$-independent $\beta_{F\vec{h}} = \beta_F$ (his eq.[6.2] and [6.3]). This contradicts our eq.(77)-(78) in the same way as Yamamoto does.

The Tanaka-Edwards model (1992a-d) rests on the same basis as Yamamoto's except that it considers a finite number of strands, with a balance between free and connected strands corresponding to that of our section 4/. They allow a free choice of $\beta_{F\vec{h}}$, with Gaussian $\Gamma_{F\vec{h}}^{eq} \propto \exp{-\chi \vec{h}^2/2k_BT}$ and constant $\beta_{C\vec{h}} = \beta_C$. Their equilibrium distribution of connected strands $\Gamma_{C\vec{h}}^{eq} = \beta_{C\vec{h}}\Gamma_{F\vec{h}}^{eq}/\beta_F$ is therefore non-Gaussian (eq.(2.23) of Tanaka-Edwards (1992$a$), corresponding to our eq.(44)). Since on the other hand these authors also choose $\zeta_{C\vec{h}} = \chi\vec{h}^2/2k_BT$ as the potential for connected strands, relation (35) is violated and the Tanaka-Edwards model suffers from the same inconsistency as Yamamoto's. (in Yamamoto's and Tanaka-Edwards' notations, the connector force $\vec{f} = \partial\phi/\partial\vec{h}$ derives from the " free energy per molecule " $\phi$. Since these authors assume that the strand volume $V_{C\vec{h}}$ does not depend on $\vec{h}$, their $\phi$ equals $\zeta_{C\vec{h}} - PV_{C\vec{h}} = \zeta_{C\vec{h}} + cst$ in our notations and their connector force $\vec{f} = \partial\phi/\partial\vec{h}$ is equivalent to our $\vec{f}_{C\vec{h}} = \partial\zeta_{C\vec{h}}/\partial\vec{h}$).

The rheometric functions of transient network models have been investigated numerically where analytical solutions are not accessible. Takano (1974) considered Yamamoto's model with Hookean strands, a disconnection rate constant $\beta_{F\vec{h}} \propto \vec{h}^2$ and a Gaussian generation function, thus violating condition (78). Among the variants of the Yamamoto model examined by Fuller and Leal (1981), the non-preaveraged version of the Phan Thien-Tanner model introduced in their eq.(33-35) uses Gaussian strands, takes $\beta_{F\vec{h}} = \beta_0(1+\sigma\vec{h}^2)$, and uses a generation function which satisfies condition (77)-(78). This model, also considered by Hermann and Petruccione (1992), is thermodynamically consistent, in contrast with the other cases examined in Fuller and Leal (1981), where condition (77)-(78) is violated. Non-Gaussian strands, with $\vec{h}$-independent $\beta_{F\vec{h}} = \beta_F$, have been considered by Vrahopoulou and McHugh (1987) in a thermodynamically consistent model. It must be noted that the previous considerations are limited to a zero value of the slip parameter $\xi$ present in Fuller and Leal (1981) and Vrahopoulou and McHugh (1987), corresponding to our affine deformation assumption. The thermodynamic intricacies introduced by the slip parameter are discussed by Larson (1983).

A last remark concerns the incidence of thermodynamic consistency of the models used in numerical computation of complex flows. Since inconsistent models are able to extract energy out of nothing, numerical instabilities then can be due to the model itself rather that being caused by a failure of the numerical scheme.

**Summary**

The present work applies the thermodynamic ideas of Prigogine and Mazur (1953) to the network models of Yamamoto (1956) and Tanaka - Edwards (1992). The formalism of chemical thermodynamics is thus applied to all changes experienced by the strands, namely the connection to the network or the disconnection from it, the variation of the connector due to the macroscopic deformation of the sample, and the Brownian diffusion of the free strands. The strand chemical potential relates the variation of the strand distribution to the entropy, energy and volume variation through the Gibbs equation. Therefore, the thermodynamic potentials keep their meaning when the system is driven out of equilibrium by thermal and/or mechanical action. Singling the part due to the deformation out of the variation of the strand distribution makes the stress naturally appear as the deformation thermodynamically conjugate quantity. This also introduces a new quantity we call the entropic stress, construed as the entropic part of the stress, relating the entropy to the deformation. The entropic stress appears in the heat equation, giving the heat generated or absorbed as a function of the thermomechanical history. The ability to cope with the thermal as well as the mechanical aspects of rheothermodynamic processes is an achievement of the present formalism.

The model is kept as simple as possible : The strands form an ideal system (in the chemical sense), the connection/disconnection reaction obeys a first order kinetics (still in the chemical sense), the Brownian diffusion of free strands is fast enough for resulting in a partial equilibrium, and heat conduction is governed by the Fourier law. The condition that the connection/disconnection reaction stops at equilibrium introduces the detailed balance condition relating the reaction rate constants to the equilibrium distribution, hence to the connector force law. The entropy source is shown to satisfy the all-important Clausius-Duhem relation : it is non-negative for all possible thermomechanical histories, and it vanishes at equilibrium. This ensures the thermodynamic consistency of our model.

The important special case of entropic strands leads to the expected identity between the extra stress and the entropic extra stress. Consideration of a constant connection rate permits to introduce the generation function used in most of the transient network literature. The detailed balance, the source of entropy and the stress are rewritten in terms of the generation function.

A review of Yamamoto's model progeny shows that all cases of thermodynamic inconsistency can be traced to the failure to obey the detailed balance condition, in the form (77) for the models expressed in terms of a generation function and in the form (45) for those explicitly considering the free strands.

**Symbols, equation of definition**

| | | | |
|---|---|---|---|
| $\alpha$ | (3) | $A$ | (20) |
| $\beta_{\alpha \vec{h}}$ | (40) | $C_V$ | (58) |
| $\Gamma_{\alpha \vec{h}}$ | (3) | $C_{V,\{\Gamma\}}$ | (18) |
| $\Gamma_{\alpha \vec{h}}^{eq}$ | (35) | $C$ (as subscript) | (3) |
| $\Delta \zeta_{\vec{h}}$ | (38) | $d^A, d^R$ | (7) |
| $\Delta \zeta$ | (66) | $d^{irr}$ | (15) |
| $\Delta \mu_{\vec{h}}$ | (47) | $E$ | (3) |
| $\zeta_\alpha$ | (63) | $\vec{f}_{\vec{h}}, f_{\vec{h} i}$ | (67) |
| $\zeta_{\alpha \vec{h}}$ | (27) | $\vec{f}_{\alpha \vec{h}}, f_{\alpha \vec{h} i}$ | (29) |
| $\theta$ | (27) | $\bar{\bar{F}}, F_{ij}$ | (4) |
| $\kappa_{ij}$ | (5) | $F$ (as subscript) | (3) |
| $\lambda$ | (46) | $g_{\vec{h}}$ | (75) |
| $\mu_{\alpha \vec{h}}$ | (3) | $\vec{h}, h_i$ | (3) |
| $\mu_{eq}$ | (33) | $k_B$ | (27) |
| $\mu_{\alpha \vec{h}}^{eq}$ | (33) | $N$ | (1) |
| $\bar{\bar{\sigma}}, \sigma_{ij}$ | (13) | $N_\alpha$ | (1) |
| $\bar{\bar{\tau}}, \tau_{ij}$ | (10) | $N_\alpha^{eq}$ | (36) |
| $\bar{\bar{\tau}}^{eq}, \tau_{ij}^{eq}$ | (34) | $P$ | (3) |
| $\phi_{\vec{h}}$ | (65) | $p(t|t',\vec{h}')$ | (52) |
| $\phi_{\alpha \vec{h}}$ | (36) | $\vec{q}$ | (14) |
| $\chi$ | (85) | $\bar{\bar{s}}_{\{\Gamma\}}, s_{\{\Gamma\}ij}$ | (22) |
| $\Omega_{\vec{h}}$ | (63) | $\bar{\bar{s}}, s_{ij}$ | (58) |
| | | $S_{\alpha \vec{h}}, S_\alpha$ | (64) |
| | | $\bar{\bar{t}}, t_{ij}$ | (24) |
| | | $S$ | (3) |
| | | $T$ | (3) |
| | | $v_{\vec{h}}$ | (40) |
| | | $V$ | (3) |
| | | $\vec{w}_{\vec{h}}, w_{\vec{h} i}$ | (42) |

**References**

Annable, T., R. Buscall, R. Ettelaie and D. Whittlestone, " The rheology of solutions of associating polymers : comparison of experimental behaviour with transient network theory ", *J. Rheol.* 34, 695-726 (1992)

Courant, R. and D. Hilbert, " Methods of mathematical physics ", Wiley (1953)

de Groot, S.R. and P. Mazur, " Non-equilibrium thermodynamics ", 2$^{nd}$ ed., Dover (1984)

Fuller, G.G. and L.G. Leal, " Network Models of Concentrated Polymer Solutions Derived from the Yamamoto Network Theory ", *J. Polym.Sci.:Polym.Phys. Ed.* 19, 531-555 (1981)

Green, M.S. and A.V.Tobolsky," A New Approach to the Theory of Relaxing Polymeric Media ", *J. Chem. Phys*. 14, 80-92 (1946)

Hermann, W. and F. Petruccione, " Quantitative rheological predictions of a transient network model of the Lodge-Yamamoto type : Simple and multiaxial elongational flow ", J. Rheol. 36, 1461-1476 (1992)

Jongschaap, R.J.J., A.I.M. Denneman, W. Conrads, « Thermodynamic approach to rheological modeling and simulation at the configuration space level of description », J. Rheol. 41, 219-235 (1997)

Landau L.D. and E.M. Lifshitz, " Statistical Physics part 1 ", §12, *Pergamon Press* (1980)

Larson R.G., « Convection and diffusion of polymer network strands », *J. Non-Newtonian Fluid Mech.*13, 279-308 (1983)

Lodge, A.S., " Constitutive Equations from Molecular Network Theories for Polymer Solutions ", *Rheol. Acta* 7, 379 (1968)

Prigogine, I. and  P. Mazur, " Sur l'extension de la thermodynamique aux phénomènes irréversibles liés aux degrés de liberté internes ", *Physica* 19, 241-254 (1953)

Sekimoto, K., " Thermodynamics and hydrodynamics of chemical gels ", *J. Physique II*, 1, 19-36 (1991)

Takano, Y., " Network Theory for Nonlinear Viscoelasticity ", *Polymer Journal* 6, 61-71 (1974)

Tanaka, F. and S.F. Edwards,
   (a) "Viscoelastic Properties of Physically Cross-Linked Networks. Transient Network Theory", *Macromolecules 25*: 1516-1523  (1992)

(b) "Viscoelastic properties of physically cross-linked networks Part 1. Non-linear viscoelasticity", *J. Non-Newtonian Fluid Mech. 43*: 247-271 (1992)

(c) "Viscoelastic properties of physically cross-linked networks Part 2. Dynamic mechanical moduli", *J. Non-Newtonian Fluid Mech. 43*: 273-288 (1992)

(d) "Viscoelastic properties of physically cross-linked networks Part 3. Time-dependent phenomena", *J. Non-Newtonian Fluid Mech. 43*: 289-309 (1992)

Vrahopoulou, E.P. and A.J. McHugh, " A consideration of the Yamamoto Network Theory with Non-Gaussian Chain Segments ", *J. Rheol. 31*, 371-384 (1987)

Yamamoto, M.J., "The Visco-elastic Properties of Network Structure I. General Formalism", *J. Phys. Soc. Jpn. 11*: 413-421 (1956)

"The Visco-elastic Properties of Network Structure II. Structural Viscosity", *J. Phys. Soc. Jpn. 12*: 1148-1158 (1957)

"The Visco-elastic Properties of Network Structure III. Normal Stress Effects (Weissenberg Effect)", *J. Phys. Soc. Jpn. 13*, 1200-1211 (1958)